# Direct reading of charge multipliers with a self-triggering CMOS analog chip with 105k pixels at 50 μm pitch

R. Bellazzini[a], G. Spandre[a], M. Minuti[a], L. Baldini[a], A. Brez[a], F. Cavalca[a],
L. Latronico[a], N. Omodei[a], M. M. Massai[b], C. Sgro'[a], E. Costa[c], P. Soffitta[c]
F. Krummenacher[d], R. de Oliveira[e]

[a]INFN sez.Pisa, Largo B. Pontecorvo, 3 I-56127 Pisa, Italy
[b]University of Pisa and INFN-Pisa, Largo B. Pontecorvo, 3 I-56127 Pisa, Italy
[c]Istituto di Astrofisica Spaziale e Fisica Cosmica of INAF, Via del Fosso del Cavaliere, 100, I-00133 Roma, Italy
[d]Ecole Polytechnique Federale de Lausanne, Lausanne, Switzerland
[e]CERN, CH-1211 Genève 23, Switzerland

**Abstract**

We report on a large active area (15x15mm$^2$), high channel density (470 pixels/mm$^2$), self-triggering CMOS analog chip that we have developed as pixelized charge collecting electrode of a Micropattern Gas Detector. This device, which represents a big step forward both in terms of size and performance, is the last version of three generations of custom ASICs of increasing complexity. The CMOS pixel array has the top metal layer patterned in a matrix of 105600 hexagonal pixels at 50μm pitch. Each pixel is directly connected to the underneath full electronics chain which has been realized in the remaining five metal and single poly-silicon layers of a standard 0.18μm digital CMOS VLSI technology. The chip has customizable self-triggering capability and includes a signal pre-processing function for the automatic localization of the event coordinates. In this way it is possible to reduce significantly the readout time and the data volume by limiting the signal output only to those pixels belonging to the region of interest. The very small pixel area and the use of a deep sub-micron CMOS technology has brought the noise down to 50 electrons ENC.
Results from in depth tests of this device when coupled to a fine pitch (50μm on a triangular pattern) Gas Electron Multiplier are presented. The matching of readout and gas amplification pitch allows getting optimal results. The application of this detector for Astronomical X-Ray Polarimetry is discussed. The experimental detector response to polarized and unpolarized X-ray radiation when working with two gas mixtures and two different photon energies is shown. Results from a full MonteCarlo simulation for several galactic and extragalactic astronomical sources are also reported.

## 1. Introduction

Since 2001 we have been actively working at INFN Pisa on the concept of Gas Pixel Detectors in which a custom CMOS analog chip is at the same time the pixelized charge collecting electrode and the amplifying, shaping and charge measuring front-end electronics of Micropattern Gas Detectors (MPGD) or other suitable charge multiplier [1]. To this purpose three ASIC generations of increasing size, reduced pitch and improved functionality have been realized (Fig.1.)

The performance obtained with the first two chips, with 2k and 22k pixels at 80μm pitch realized in 0.35μm technology, and their response to polarized and unpolarized X-ray radiation has been discussed elsewhere [2,3]. In this paper we discuss the results achieved with the last ASIC version we have designed in 0.18μm VLSI technology.

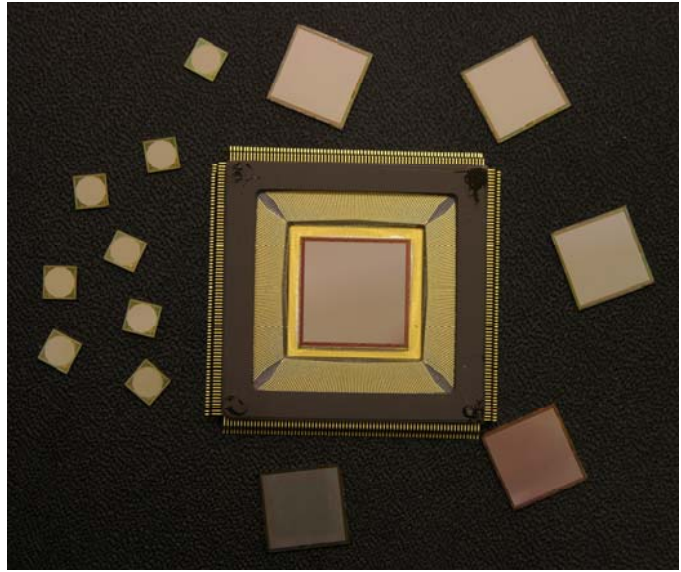

**Fig.1 – The three chip generations in comparison.
The last version with 105.600 pixels is shown bonded to its ceramic package (304 pins).**

## 2. The third CMOS VLSI generation and the MPGD

The chip has 105600 hexagonal pixels arranged at 50μm pitch in a 300×352 honeycomb matrix, corresponding to an active area of 15×15mm$^2$ with a pixel density of 470/mm$^2$. Each pixel is connected to a charge-sensitive amplifier followed by a shaping circuit and a sample&hold-multiplexer circuit. The chip integrates more than 16.5 million transistors and it is subdivided in 16 identical clusters of 6600 pixels (22 rows of 300 pixels) or alternatively in 8 clusters of 13200 pixels (44 rows of 300 pixels) each one with an independent differential analog output buffer (Fig.2).

Each cluster has a customizable internal self-triggering capability with independently adjustable thresholds. Every 4 pixels (mini-cluster, Fig.3) contribute to a local trigger with a dedicated amplifier whose shaping time (Tshaping~1.5 μs) is roughly a factor two faster than the shaping time of the analog charge signal. The contribution of any pixel to the trigger can be disabled by direct addressing of the pixel. An internal wired-OR combination of each mini-cluster self-triggering circuit holds the maximum of the shaped signal on each pixel. The event is localized in a rectangular area containing all triggered miniclusters plus a user selectable margin of 10 or 20 pixels. The Xmin, Xmax and Ymin, Ymax rectangle coordinates are available as four 9-bit data outputs as soon as the data acquisition process following an internally triggered event has terminated, flagged by the DataReady output. The event window coordinates can be copied into a Serial-Parallel IO interface register (a

36-stage FIFO) by applying an external command signal (ReadMinMax). Subsequently, clock pulses push out the analog data to a serial balanced output buffer compatible with the input stage of the Texas Instruments 12 bit flash ADC ADS572x.

In self-trigger operation the read-out time and the amount of data to be transferred result vastly reduced (at least a factor 100) with respect to the standard sequential read-out mode of the full matrix (which is still available, anyway). This is due to the relatively small number of pixels (600-700) within the region of interest.

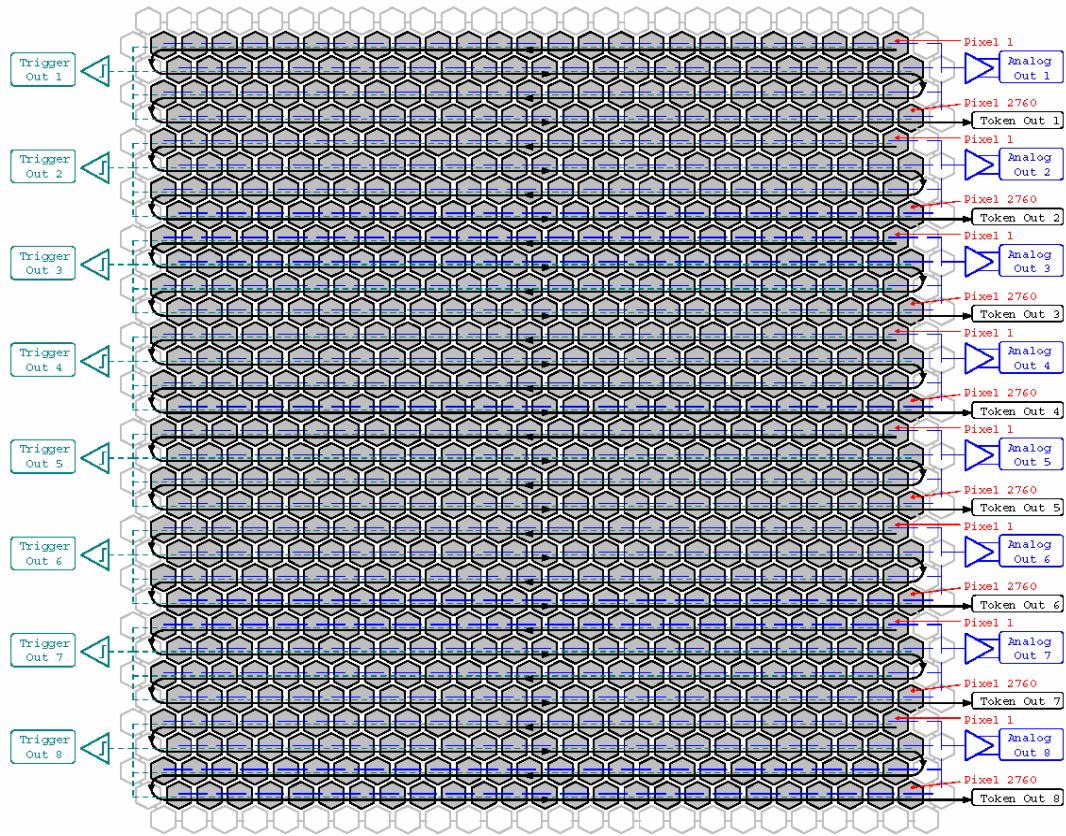

**Fig. 2 – Simplified pixel layout.**

Main characteristics of the chip are:
- Peaking time: 3-10 μs, externally adjustable;
- Full-scale linear range: 30000 electrons;
- Pixel noise: 50 electrons ENC;
- Read-out mode: asynchronous or synchronous;
- Trigger mode: internal, external or self-trigger;
- Read-out clock: up to 10MHz;
- Self-trigger threshold: 2200 electrons (10% FS);
- Frame rate: up to 1 kHz in self-trigger mode (event window);
- Parallel analog output buffers: 1, 8 or 16;
- Access to pixel content: direct (single pixel) or serial (8-16 clusters, full matrix, region of interest);
- Fill fraction (ratio of metal area to active area): 92%

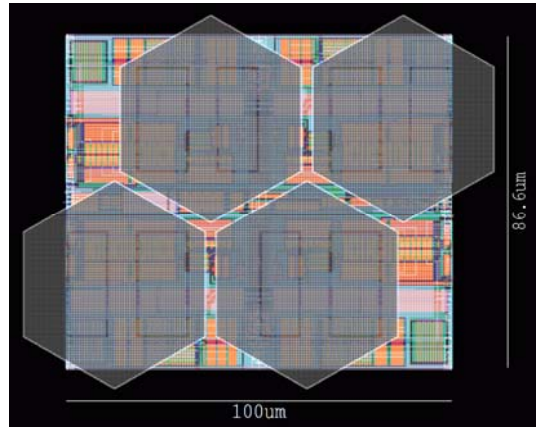

Fig. 3 – The 4 pixel self-trigger mini-cluster definition.

A 50 μm thick Gas Electron Multiplier with 50 μm pitch holes on a triangular pattern has been assembled on top of chip. The perfect matching of read-out and gas amplification sampling allows to get optimal results and to fully exploit the very high granularity of the device. The technological challenge in the fabrication of this type of GEM was the precise and uniform etching of very narrow charge multiplication holes (30 and 23μm diameter at the top and in the middle of the kapton layer) on a GEM foil of standard 50 μm thickness. The *absorption region* of the Gas Pixel Detector between the GEM and the enclosure drift electrode (25μm Aluminized Mylar foil), has been realized with a 10 mm spacer. The *collection gap* between the bottom GEM and the pixel matrix of the read-out chip is roughly 1 mm thick. The detector response has been studied with two different gas mixtures: our standard 50% Neon – 50% DME mixture and a *lighter* one, composed by 40% Helium – 60% DME. Typical high voltage settlings used in the tests for the Neon based mixture are: $V_{DRIFT}$ = -1800V, $V_{GEM}(top)$ = -750V, $V_{GEM}(bottom)$ = -300V, the read-out electrode being at ~0V (the charge preamp input voltage). The He based mixture reaches the same gas gain at $V_{GEM}(top)$ = -780V, essentially because of the larger DME content. The 50 μm pitch GEM has shown to work magnificently with both gas mixtures. It has a large effective gain (well above 1000) at a much reduced voltage (at least 70 Volt less) comparing with our previous 90 μm pitch GEM. Likely, this is due to the higher field lines density inside the very narrow amplification holes. A photo of the detector ready to be mounted on the control motherboard is shown in Fig.4.

A custom and very compact DAQ system to generate and handle command signals to/from the chip (implemented on Altera FPGA Cyclone EP1C240), to read and digitally convert the analog data (ADS5270TI Flash ADC) and to storage them, temporarily, on a static RAM, has been developed. By using the RISC processor NIOS II, embedded on Altera FPGA, and the self-triggering functionality of the chip, it is possible, immediately after the event is read-out, to acquire the pedestals of the pixels in the same *chip-defined* event window (region of interest). The pedestals can be read one or more times (user-defined). The average of the pedestal readings is used to transfer pedestal subtracted data to the off-line analysis system. This mode of operation has the great advantage of allowing the real time control of the data quality and to cancel any effect of time drift of the pedestal values or other temperature or environmental effects.

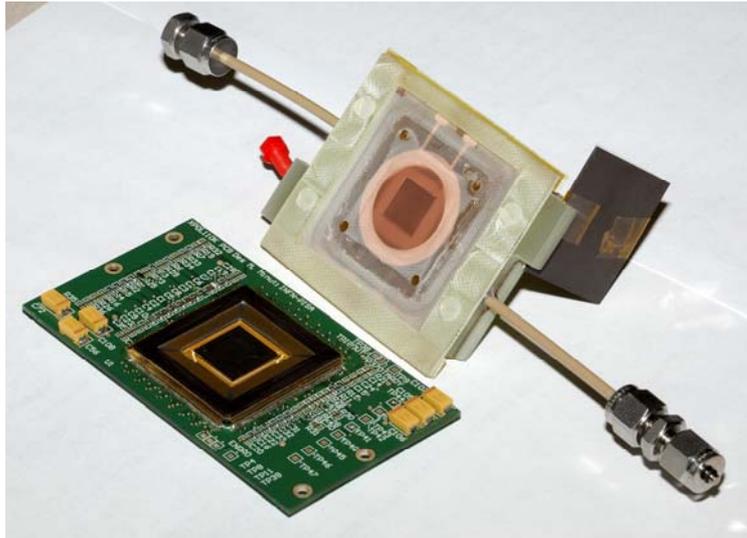

**Fig. 4 – Photo during the assembly phase of the detector.
The GEM foil glued to bottom of the gas-tight enclosure and the large area ASIC mounted on the control motherboard are well visible.**

The side disadvantages are a slight increase of the channel noise (at maximum a factor $\sqrt{2}$, for the case of 1 pedestal reading only) and an increase of the event read-out time. For most of the applications we envisage, this is not a real problem giving the very large signal to noise ratio (well above 100) and the very fast window mode operation. Anyway, the standard mode of operation with the acquisition of a set of pedestal values for all the 105k channels at the beginning or at the end of a data taking run is still possible. The instrument control and data acquisition is done through a VI graphic interface developed in LAbVIEW. The VI performs a bidirectional communication through a 100Mbps TCP connection between the DAQ and a portable PC running Windows XP.
Fig.5 shows a snapshot of the Control Panel of the VI.

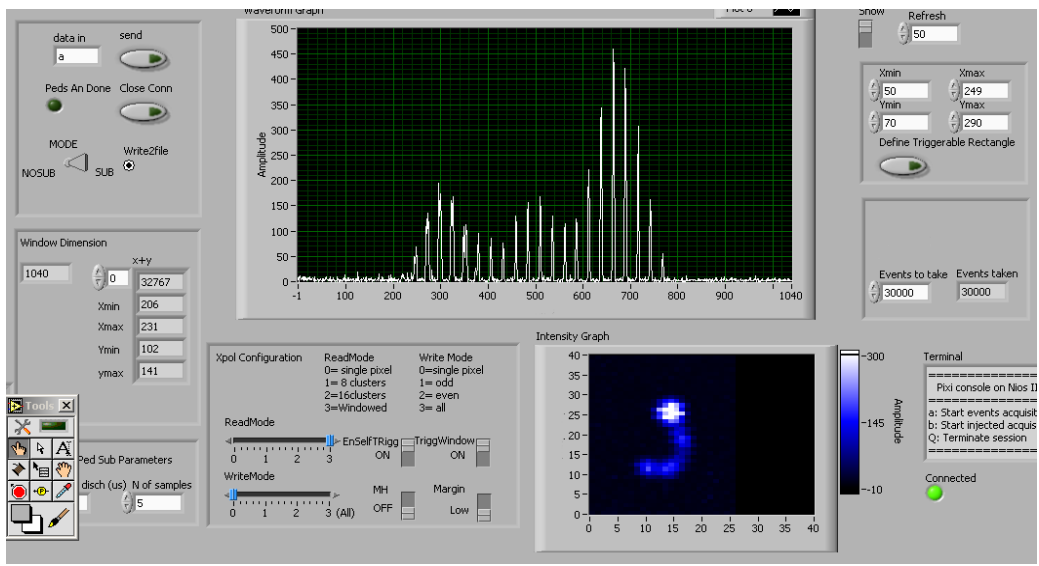

**Fig. 5 – Snapshot of the Control Panel of the acquisition system. Together with the sets of user-selectable commands (chip configuration, self-trigger enable, on board pedestal subtraction,..) the canvas shows the display of one event as one-dimensional pixel charge distribution (ADC counts vs. pixel number in the event window) and the relative 2D image displayed in real time.**

## 3. Tests on the chip

A complete set of tests has been performed on the chip. Fig.6 shows the noise distribution for all the 105600 pixels measured as rms value of the pedestal fluctuations; the average noise level for the entire chip is around 3 ADC counts corresponding to ~50 electrons ENC (amplifier input sensitivity = 350 ADC counts/fC).

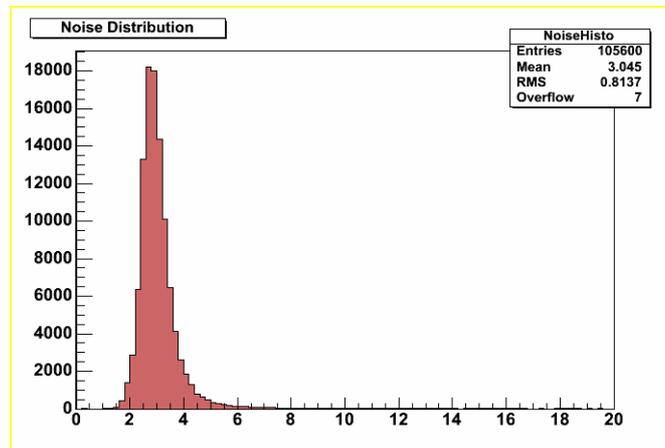

**Fig. 6 - Noise distribution for all the 105k pixels. Amplifier input sensitivity = 350 ADC counts/fC. All pixels are working.**

The most interesting feature of the chip is, undoubtedly, its capability to operate in self-trigger mode. Once the trigger thresholds are set (we use just a global one for the whole chip) and the self-trigger enabled in the configuration register, the chip is ready to localize automatically the region (*Event Window*) around the pixels which have contributed to the trigger and to start the peak search and hold of the shaped signals of these pixels. The Event Window is a rectangle that contains all triggered mini-clusters plus a user-selectable margin of 10 or 20 pixels (0.5mm or 1.0mm). These margins allow recovering the information carried by the pixels whose charge content is below the global threshold but above the individual threshold which can be applied off-line (typically 4 sigmas of the channel noise, i.e. 200 electrons).

The signals shown in Fig.7 refer to the chip working in self-trigger mode. Upon the activation of the *Write* signal, a calibration charge of 1 fC is injected into the preamplifier input (the blue track indicates the analog pulse on the output buffer) generating a signal higher than the global trigger threshold, actually set to 0.4 fC. The trigger circuit and logic inside the chip generates, within ~1 µs, the *TriggerOut* signal (orange track) which starts the peak detection of the shaped analog pulse. At the end of this process (~ 7 µs later) the *DataRready* signal (green track) is generated.

The lowest applicable global trigger threshold is constrained by pedestal offsets more than pedestal fluctuations. The offset variation range in CMOS technology is typical 10% of the linear dynamics.

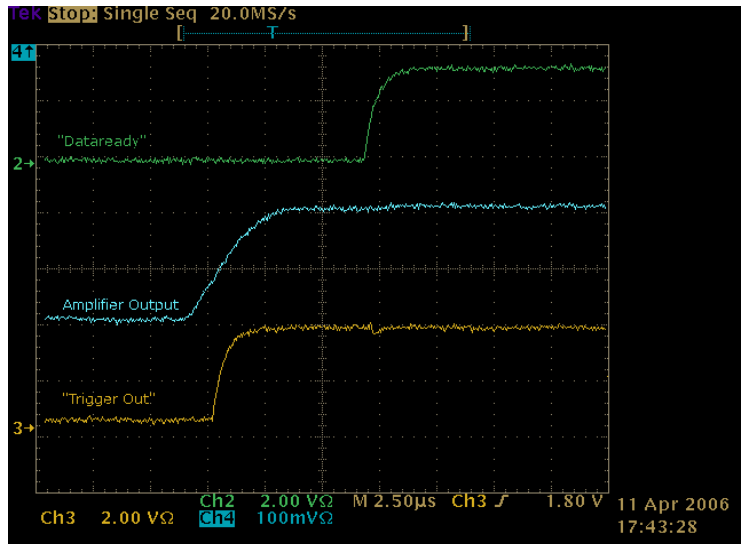

**Fig. 7 – Self-trigger mode and timing: a 1 fC charge, above the trigger global threshold (0.4 fC), is injected into the preamp input (the blue track indicates the analog pulse on the output buffer); the trigger circuit generates after ~1μs the *TriggerOut* signal (orange track) which starts the search of the signal maximum. The *DataReady* signal is generated at the end of this process (green track).**

By varying the threshold the fake trigger rate has been measured in self-trigger functionality (Fig.8). All the pixels were working and no mask was applied to kill noisy pixels. Our typical threshold working point was then set to ~2300 electrons where the fake trigger rate is reduced to 3 Hz or less. This threshold is much lower of the one we were able to apply when working in external trigger mode using the top GEM signal to start the read-out sequence

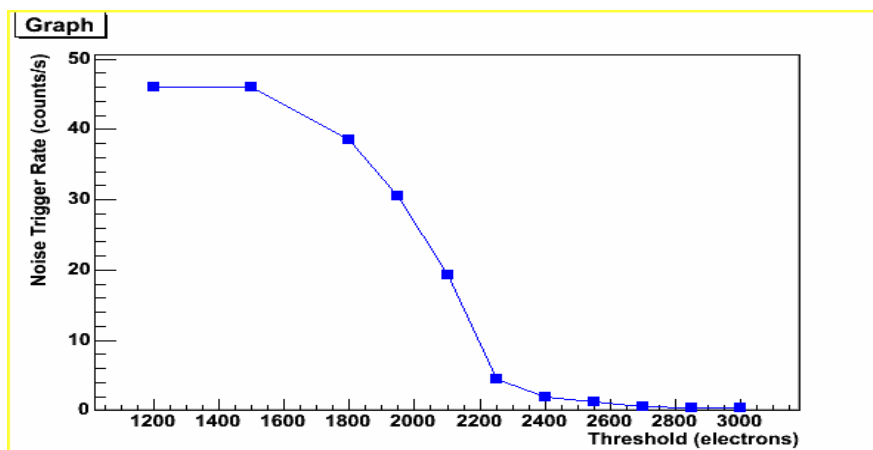

**Fig. 8 – Fake trigger rate vs. threshold.**

The chip has been tested with X-rays from a $^{55}$Fe source (5.9 keV) and an X-ray generator with Cr anode (5.41 keV) always working in self-trigger mode. On average, the size of the region of interest around the triggering event is ~700 pixels (see Fig.9).

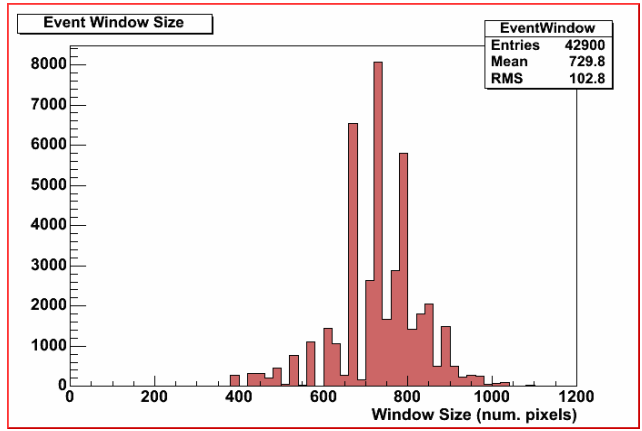

**Fig. 9 – Event window size distribution.
On average only 730 pixels, over 105600 totals, are read.**

## 4. Performance of the MPGD

The use of a GEM with a much finer pitch than usual (pitch and thickness have now equal size) and well matching the 50μm read-out pitch has pushed forward the 2D imaging capability of the device, allowing to reach a very high degree of detail in the photo-electron track reconstruction. This is of great importance for our X-Ray Polarimetry application [2], especially when working at low photon energy (2-3 keV). Some *real track* events are shown in Fig.10.

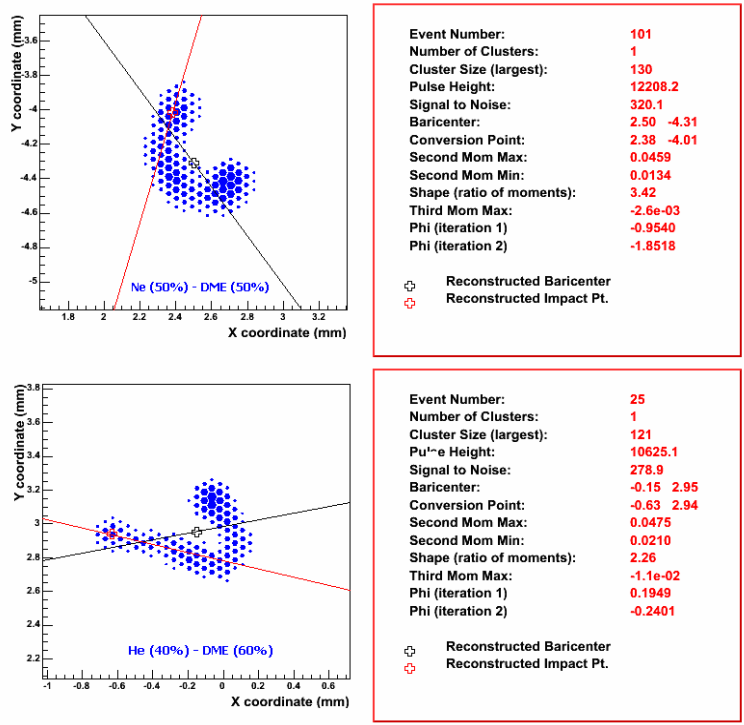

**Fig. 10 – Real tracks obtained by irradiating the detector with X-rays from the Cr tube for two gas mixtures. The black cross refers to the barycenter position and the black line to the direction of the principal axis (iteration 1); the red cross refers to the conversion point evaluation and the red line to the emission direction of the photo-electron (iteration 2)**

A fine grain reconstruction of the initial part of the photo-electron track allows a better estimation of the emission direction and from the obtained angular distribution a better evaluation of the degree and angle of polarization of the detected radiation.

Fig.11 shows the cumulative hit map of 100 tracks obtained with photons from a Cr X-ray tube using a gas filling of 40% Helium and 60% DME. This figure well explains how the detector works as X-Ray Polarimeter. On an event by event basis, and without any rotation of the set-up, the direction of emission of the photoelectron can be reconstructed and the modulation of corresponding angular distribution measured.

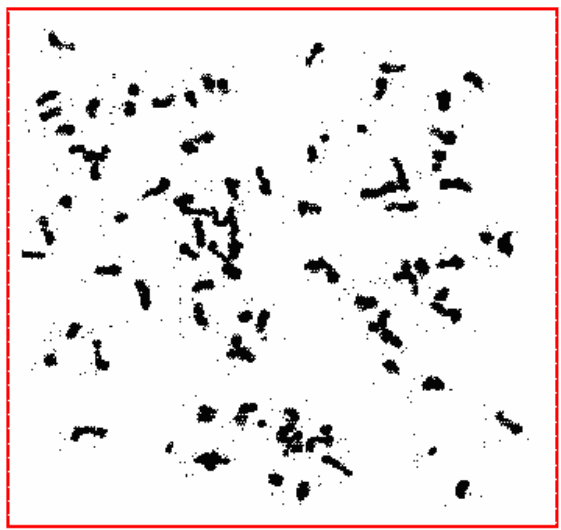

**Fig. 11 – Cumulative hit map of 100 photo-electron tracks in He (40%)-DME (60%). Photons from Cr X-ray generator.**

The very good imaging capability of the detector has been tested by illuminating with a $^{55}$Fe source, from above, a small pendant (few mm in size) placed in front of the detector. The *radiographic* image is obtained by plotting both the barycenters and the conversion points (Fig.12). It is worth to note how, for Neon or He based mixtures the accuracy of image reconstruction is much better using the absorption point instead of the barycenter. In these *light* mixtures, where tracks are long with respect to the detector granularity, the barycenter can be quite far away from the point of photon absorption. This effect is well evident in Fig. 13, where a reconstructed multi-holes phantom is shown.

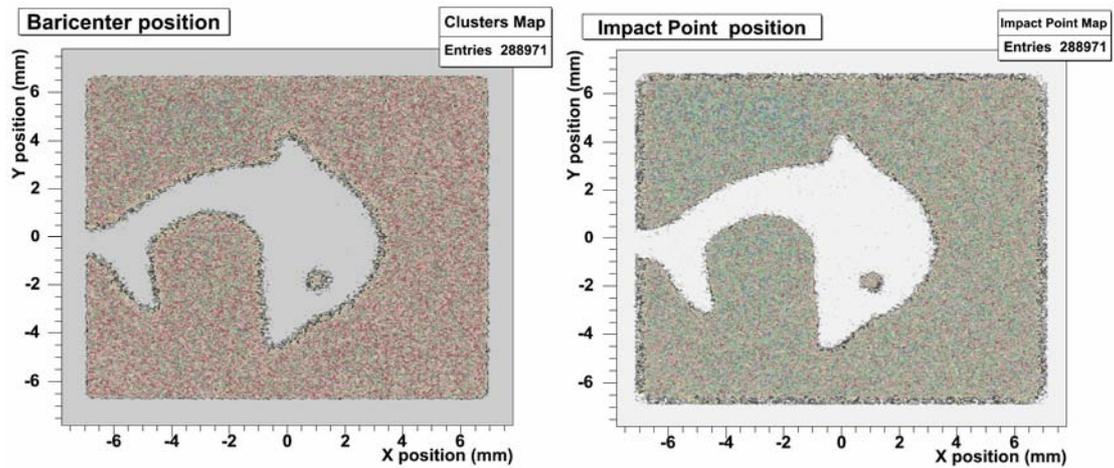

**Fig. 12 -** *Radiographic* image of a small pendant obtained with photons from a 55Fe source.

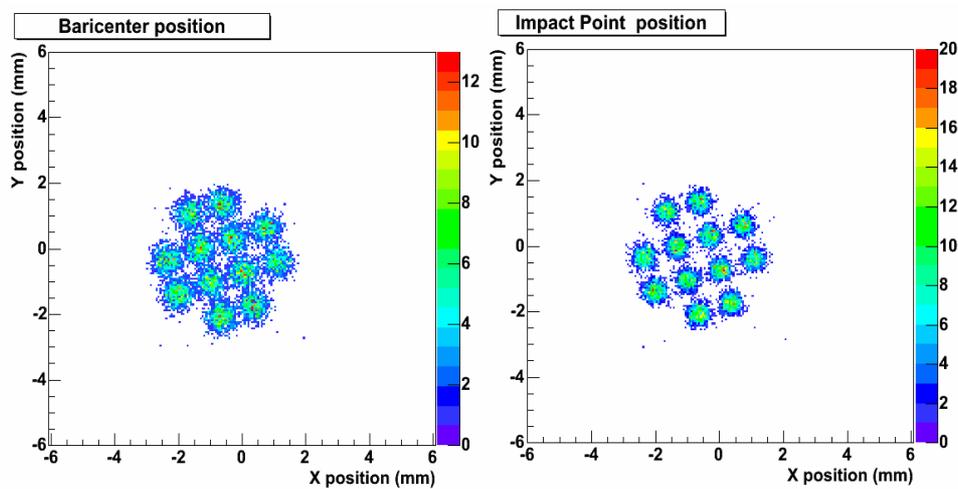

**Fig. 13 -** Differences in image reconstruction using barycenters or conversion points (impact point in figure) in Ne (50%)-DME(50%). Holes: 0.6 mm diameter, 2 mm apart.

## 5. Polarization measurements

Measurement of cosmic X-ray polarization can shed light on the structure of very compact sources and derive information on mass and angular momentum of supermassive objects. For this reason the expectations for science return from X-Ray polarimetry are large within the high-energy astrophysics community. The possibility to reconstruct photoelectron tracks with the degree of detail shown by the device presented here makes this detector ideal candidate as high sensitivity photoelectric polarimeter. To characterize the detector as X-Ray polarimeter, first of all we have measured the value of the residual modulation due to systematic errors.

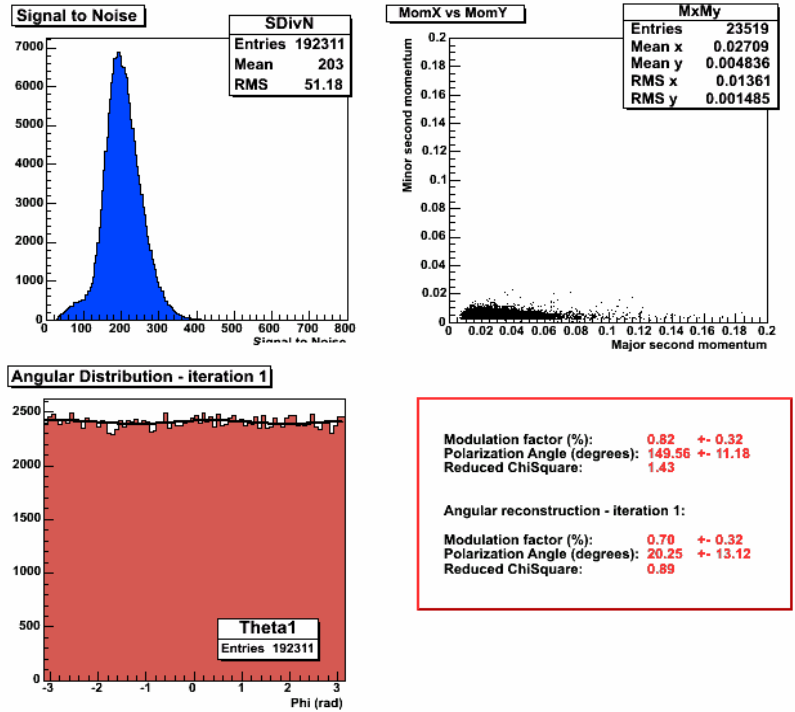

Fig. 14 – S/N distribution, scatter plot of the two principal axes of the cluster charge and residual modulation, obtained with $^{55}$Fe source with Ne(50%)-DME(50%) gas mixture filling and $\Delta V_{GEM} \sim 450V$.

The residual modulation, if any, sets the limit on the Minimum Detectable Polarization (MDP). With totally unpolarized photons from a $^{55}$Fe X-ray source a modulation of 0.70% ± 0.32% (statistical error) has been measured (Fig.14, bottom-left panel). Data have been taken with a gas filling of Ne(50%)-DME(50%) and a voltage difference through the GEM of 450V.

In these conditions we have obtained an average S/N ratio of 203 (Fig.14, top-left panel) with an average cluster size of 90 pixels. Fig.14 (top-right panel) shows also the scatter plot of the two principal axes of the cluster charge distribution. The elongated shape of the tracks is well evident. This fact is crucial for the precise determination of the emission direction of the photoelectron.

The measurement of the modulation factor for polarized photons has been carried out by using radiation from a Cr X-ray tube (5.41 keV line). The X-ray beam is Thompson scattered through a Li target (6mm in diameter, 70mm long), canned in a beryllium case (500 μm thick) in order to prevent oxidation and nitridation from air [4]. The geometry of the output window of the scatterer and the distance with respect to the detector limit the scattering angles to ~90º so that the radiation impinging the detector is highly linearly polarized (better than 98%). In the first set of data, we have noticed in the cumulative hit map the presence of two well localized, nearly monochromatic (Fig.15) and highly polarized (Fig.16) spots which could be completely cut off by a Vanadium filter. Vanadium has a K-edge at 5.46 keV, so it strongly absorbs photons with energies above this edge while it is relatively transparent to the Cr fluorescence line at 5.41 keV. A possible source of these spots could be the fluorescence emission of some element within the scatterer box (likely Iron, which emits at 6.4 keV) excited by the long *bremsstrahlung* tail of the X-ray tube (operated at 20kV).

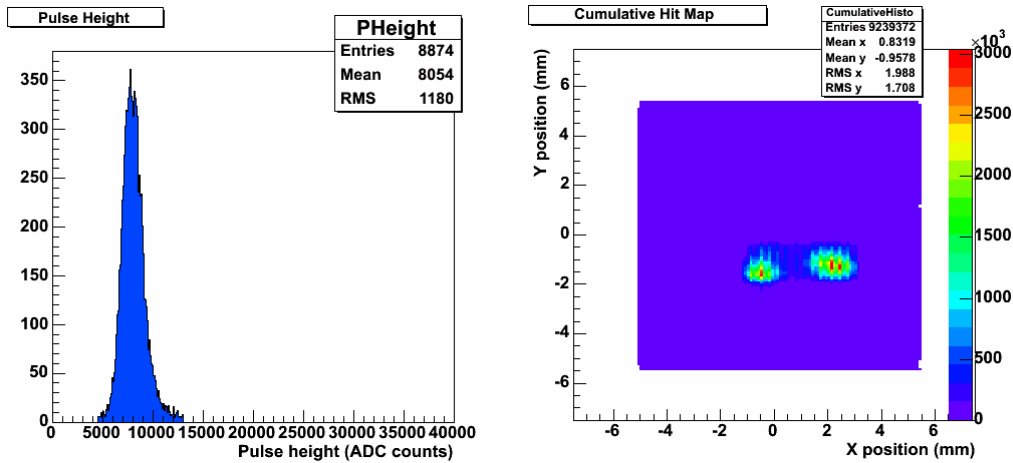

**Fig. 15 – The events belonging to the two spots in the cumulative hit map of the right panel show a very narrow pulse height distribution at ~6.4 keV**

By selecting events belonging to these *hot* spots a modulation factor of 57.01% ± 1.38% has been measured in Ne(50%)-DME(50%) (Fig.16, left panel). As expected, a even better result is obtained in the *lighter* mixture of He(40%)-DME(60%) where photoelectron tracks are longer and Auger electron tracks shorter ($E_{Auger}$~0.2 keV). In this case the modulation factor is 64.33% ± 1.54% (Fig.16, right panel).

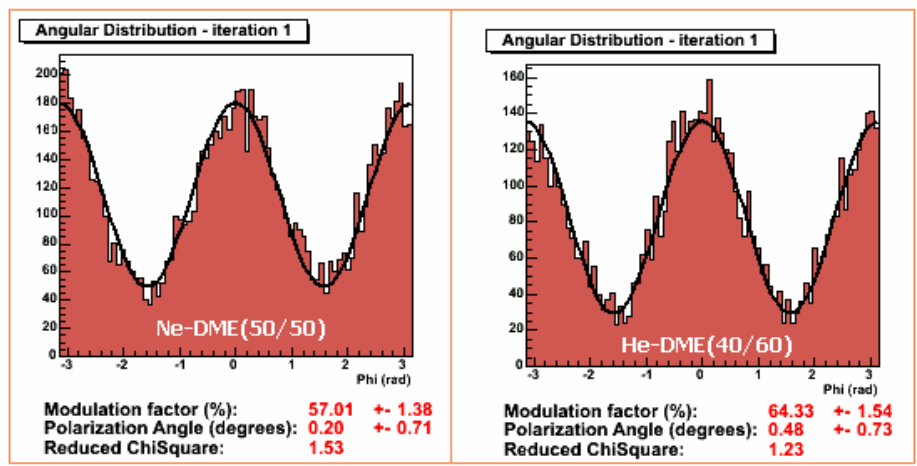

**Fig. 16 – Modulation factor measured in two different gas mixtures by selecting only the events in the hot spots (see Fig.17) at 6.4 KeV**

With the Vanadium filter in front of the entrance window of the detector (no spots in the hit map, left panel in Fig.17) the modulation factor in Ne(50%)-DME(50%) at the Cr-line energy is 51.11% ± 0.89 % and in He(40%)-DME(60%) is 54.26% ± 1.24.

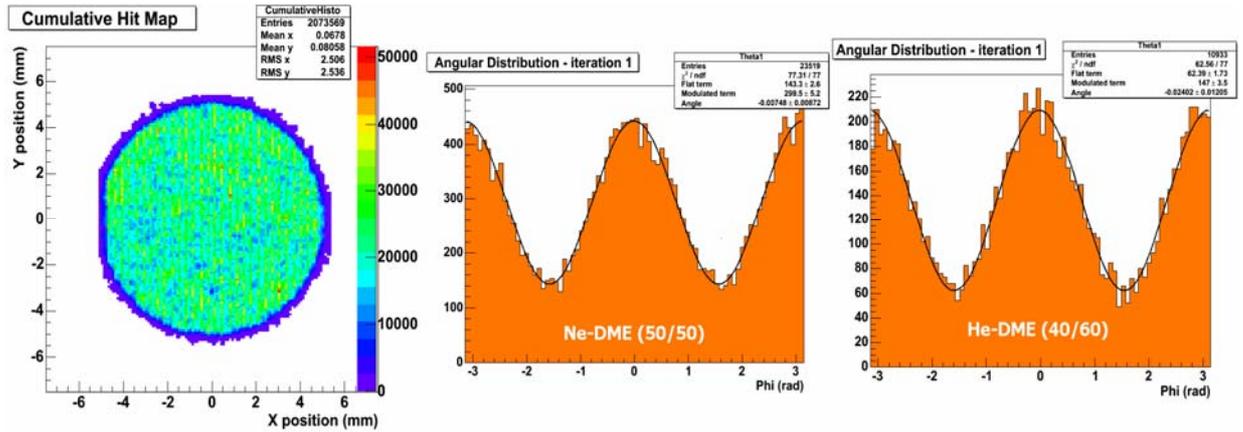

**Fig. 17 – Modulation factor in two different gas mixtures at 5.4 keV Cr-line energy (center and right panel). The Vanadium filter on the entrance window of the detector absorbs photons with energy higher than the Cr-line, removing the *hot* spots in the cumulative map (left panel).**

## 6. MonteCarlo predictions

A full MonteCarlo simulation which takes into accounts all the physics processes which rule the operation of this detector as X-ray polarimeter has been developed. These processes include photoelectric interaction, scattering and slowing of the primary electrons in the gas, drift and diffusion, gas multiplication and the final charge collection on the read-out plane. All of them are function of the photon energy and of gas parameters such as composition, pressure and drift path. A description of the computational model can be found in references [5,6].

The polarimetric sensitivity of the detector described in this paper at the focus of the XEUS optics (an ESA permanent space borne X-ray observatory planned to be launched in 2015) has been studied for a set of representative X-ray sources. The obtained result in terms of Minimum Detectable Polarization (integrated in the energy range 2-10 keV) for 1cm – 1atm of He(40%) – DME(60%) gas mixture is shown in Fig. 18.

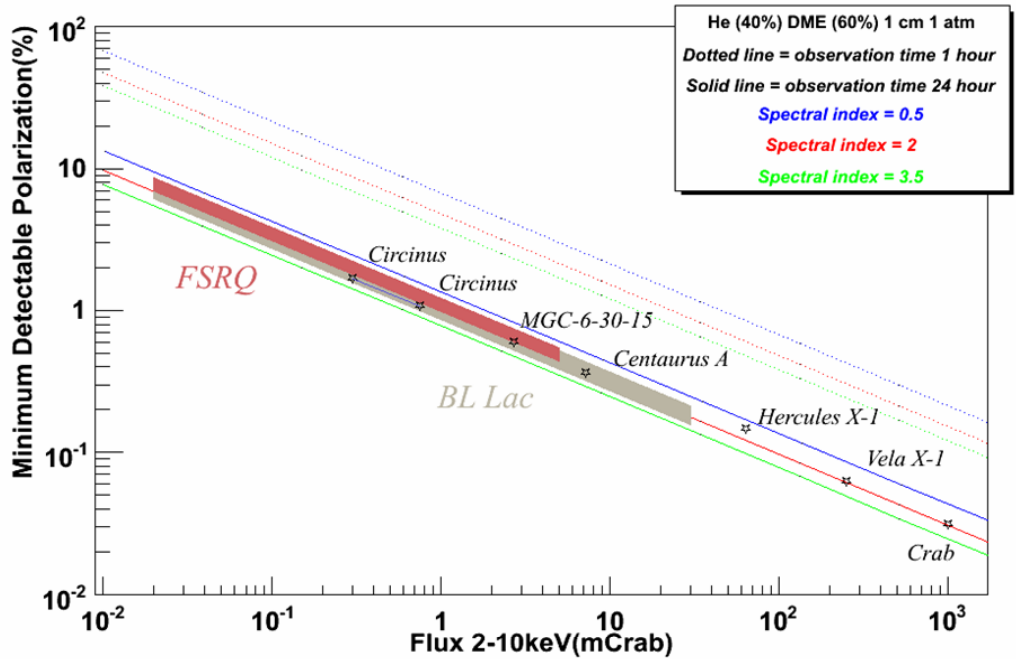

**Fig. 18 – Minimum Detectable Polarization for the detector described in the paper and proposed as Polarimeter for the XEUS optics for a few X-ray sources with milliCrab fluxes.**

With observations of one day we can measure the polarization of several AGNs down to few % levels. For its high sensitivity this detector has been proposed at the focus of a large area telescope as those ones foreseen for the New Generation X-ray Telescope in the frame of the ESA Cosmic Vision 2015-2025.

**Conclusions**

With devices like the one described in this paper the class of Gas Pixel Detectors has reached a level of integration, compactness and resolving power so far considered in the reach of solid state detectors only. As for the X-Ray Polarimetry application, the very low residual modulation and a modulation factor well above 50% will allow polarimetric measurements at the level of ~1% for hundreds of galactic and extragalactic sources: a real breakthrough in X-ray astronomy.
It must be underlined also that, depending on type of electron multiplier, pixel and die size, electronics shaping time, analog vs. digital read-out, counting vs. integrating mode, many other applications can be envisaged for this class of detectors. As an example, we are now testing in the lab a UV photodetector with a semitransparent CsI photocathode coupled to this chip. The device has single electron sensitivity and good imaging capability. Results will be reported soon.
It is also worth noticing that following a similar approach, a digital counting chip developed for medical applications (Medipix2) has been shown to work when coupled to GEM or Micromegas gas amplifiers for TPC application at the next generation of particle accelerators [7,8].
Last but not least, it is interesting to note how after several months of intensive operation, no die or single pixel have been lost for electrostatic or GEM discharges, or for any other reason.